\pdfoutput=1
\documentclass[12pt,a4paper]{article}

\usepackage{ifthen} 
\usepackage{booktabs}

\newboolean{pdflatex}
\setboolean{pdflatex}{true} 

\newboolean{articletitles}
\setboolean{articletitles}{true} 

\newboolean{uprightparticles}
\setboolean{uprightparticles}{false} 

\def\paperauthors{LHCb collaboration} 
\def\paperasciititle{Curvature-bias corrections using a pseudomass method} 
\def\papertitle{Curvature-bias corrections\\ using a pseudomass method} 
\def\paperkeywords{{High Energy Physics}, {LHCb}} 
\def\papercopyright{\the\year\ CERN for the benefit of the LHCb collaboration} 
\def\paperlicence{CC BY 4.0 licence}
\def\paperlicenceurl{https://creativecommons.org/licenses/by/4.0/}

\def\figwidth{0.49\textwidth}

\usepackage[top=1in, bottom=1.25in, left=1in, right=1in]{geometry}

\usepackage{microtype}
\usepackage{lineno}  
\usepackage{xspace} 
\usepackage{caption} 

\usepackage{graphicx}  
\usepackage{color}
\usepackage{colortbl}
\graphicspath{{./figs/}} 

\usepackage{amsmath} 
\usepackage{amssymb}
\usepackage{amsfonts}
\usepackage{upgreek} 

\newcommand*\patchAmsMathEnvironmentForLineno[1]{%
\expandafter\let\csname old#1\expandafter\endcsname\csname #1\endcsname
\expandafter\let\csname oldend#1\expandafter\endcsname\csname
end#1\endcsname
 \renewenvironment{#1}%
   {\linenomath\csname old#1\endcsname}%
   {\csname oldend#1\endcsname\endlinenomath}%
}
\newcommand*\patchBothAmsMathEnvironmentsForLineno[1]{%
  \patchAmsMathEnvironmentForLineno{#1}%
  \patchAmsMathEnvironmentForLineno{#1*}%
}
\AtBeginDocument{%
\patchBothAmsMathEnvironmentsForLineno{equation}%
\patchBothAmsMathEnvironmentsForLineno{align}%
\patchBothAmsMathEnvironmentsForLineno{flalign}%
\patchBothAmsMathEnvironmentsForLineno{alignat}%
\patchBothAmsMathEnvironmentsForLineno{gather}%
\patchBothAmsMathEnvironmentsForLineno{multline}%
\patchBothAmsMathEnvironmentsForLineno{eqnarray}%
}

\usepackage{hyperxmp}

\usepackage[pdftex,
            pdfauthor={\paperauthors},
            pdftitle={\paperasciititle},
            pdfkeywords={\paperkeywords},
            pdfcopyright={Copyright (C) \papercopyright},
            pdflicenseurl={\paperlicenceurl}]{hyperref}

\usepackage[colorinlistoftodos,textsize=scriptsize]{todonotes}

\usepackage[bottom,flushmargin,hang,multiple]{footmisc}

\usepackage[all]{hypcap} 

\usepackage{xspace} 
\usepackage{upgreek}

\def\lhcb   {\mbox{LHCb}\xspace}

\def\MagUp {\mbox{\em Mag\kern -0.05em Up}\xspace}

\ifthenelse{\boolean{uprightparticles}}%
{

 \def\Pmu         {\ensuremath{\upmu}\xspace}

 \def\Ppsi        {\ensuremath{\uppsi}\xspace}

 \def\PDelta      {\ensuremath{\Delta}\xspace}                 
 \def\PXi         {\ensuremath{\Xi}\xspace}                 
 \def\PLambda     {\ensuremath{\Lambda}\xspace}                 
 \def\PSigma      {\ensuremath{\Sigma}\xspace}                 
 \def\POmega      {\ensuremath{\Omega}\xspace}                 
 \def\PUpsilon    {\ensuremath{\Upsilon}\xspace}
 \let\oldPi\Pi
 \def\PPi         {\ensuremath{\oldPi}\xspace}

 \def\PB      {\ensuremath{\mathrm{B}}\xspace}                 
                  
 \def\PD      {\ensuremath{\mathrm{D}}\xspace}

 \def\PJ      {\ensuremath{\mathrm{J}}\xspace}                 
 \def\PK      {\ensuremath{\mathrm{K}}\xspace}

 \def\PW      {\ensuremath{\mathrm{W}}\xspace}

 \def\PZ      {\ensuremath{\mathrm{Z}}\xspace}

 \def\Pi      {\ensuremath{\mathrm{i}}\xspace}

 \def\Ps      {\ensuremath{\mathrm{s}}\xspace}

 \def\thebaroffset{0.0em}
}
{

 \def\Pmu         {\ensuremath{\mu}\xspace}

 \def\Ppsi        {\ensuremath{\psi}\xspace}                 
                  
 \mathchardef\PDelta="7101
 \mathchardef\PXi="7104
 \mathchardef\PLambda="7103
 \mathchardef\PSigma="7106
 \mathchardef\POmega="710A
 \mathchardef\PUpsilon="7107
 \mathchardef\PPi="7105
                  
 \def\PB      {\ensuremath{B}\xspace}                 
                  
 \def\PD      {\ensuremath{D}\xspace}

 \def\PJ      {\ensuremath{J}\xspace}                 
 \def\PK      {\ensuremath{K}\xspace}

 \def\PW      {\ensuremath{W}\xspace}

 \def\PZ      {\ensuremath{Z}\xspace}

 \def\Pi      {\ensuremath{i}\xspace}

 \def\Ps      {\ensuremath{s}\xspace}

 \def\thebaroffset{0.18em}
}
\newcommand{\offsetoverline}[2][\thebaroffset]{\kern #1\overline{\kern -#1 #2}}%

\makeatletter
\ifcase \@ptsize \relax
  \newcommand{\miniscule}{\@setfontsize\miniscule{4}{5}}
\or
  \newcommand{\miniscule}{\@setfontsize\miniscule{5}{6}}
\or
  \newcommand{\miniscule}{\@setfontsize\miniscule{5}{6}}
\fi
\makeatother

\DeclareRobustCommand{\optbar}[1]{\shortstack{{\miniscule (\rule[.5ex]{1.25em}{.18mm})}
  \\ [-.7ex] $#1$}}

\def\mup        {{\ensuremath{\Pmu^+}}\xspace}
\def\mun        {{\ensuremath{\Pmu^-}}\xspace} 

\def\squark    {{\ensuremath{\Ps}}\xspace}

\def\KorKbar {\kern \thebaroffset\optbar{\kern -\thebaroffset \PK}{}\xspace}

\def\D       {{\ensuremath{\PD}}\xspace}

\def\DorDbar {\kern \thebaroffset\optbar{\kern -\thebaroffset \PD}\xspace}
\def\Dz      {{\ensuremath{\D^0}}\xspace}

\def\Dp      {{\ensuremath{\D^+}}\xspace}
\def\Dm      {{\ensuremath{\D^-}}\xspace}

\def\DpDm    {\ensuremath{\Dp {\kern -0.16em \Dm}}\xspace}

\def\B       {{\ensuremath{\PB}}\xspace}

\def\BorBbar {\kern \thebaroffset\optbar{\kern -\thebaroffset \PB}\xspace}

\def\Bd      {{\ensuremath{\B^0}}\xspace}

\def\BdorBdbar {\kern \thebaroffset\optbar{\kern -\thebaroffset \Bd}\xspace}

\def\Bs      {{\ensuremath{\B^0_\squark}}\xspace}

\def\BsorBsbar {\kern \thebaroffset\optbar{\kern -\thebaroffset \Bs}\xspace}

\def\jpsi     {{\ensuremath{{\PJ\mskip -3mu/\mskip -2mu\Ppsi}}}\xspace}

\def\Y#1S{\ensuremath{\PUpsilon{(#1S)}}\xspace}

\def\LorLbar     {\kern \thebaroffset\optbar{\kern -\thebaroffset \PLambda}\xspace}

\def\to                 {\ensuremath{\rightarrow}\xspace}

\def\AT#1     {\ensuremath{A_{\mathrm{T}}^{#1}}\xspace}           

\def\C#1      {\ensuremath{\mathcal{C}_{#1}}\xspace}                       
\def\Cp#1     {\ensuremath{\mathcal{C}_{#1}^{'}}\xspace}                    
\def\Ceff#1   {\ensuremath{\mathcal{C}_{#1}^{\mathrm{(eff)}}}\xspace}        
\def\Cpeff#1  {\ensuremath{\mathcal{C}_{#1}^{'\mathrm{(eff)}}}\xspace}       
\def\Ope#1    {\ensuremath{\mathcal{O}_{#1}}\xspace}                       
\def\Opep#1   {\ensuremath{\mathcal{O}_{#1}^{'}}\xspace}                    

\newcommand{\nospaceunit}[1]{\ensuremath{\text{#1}}}       
\newcommand{\aunit}[1]{\ensuremath{\text{\,#1}}}       

\newcommand{\tev}{\aunit{Te\kern -0.1em V}\xspace}
\newcommand{\gev}{\aunit{Ge\kern -0.1em V}\xspace}
\newcommand{\mev}{\aunit{Me\kern -0.1em V}\xspace}
\newcommand{\kev}{\aunit{ke\kern -0.1em V}\xspace}
\newcommand{\ev}{\aunit{e\kern -0.1em V}\xspace}
 
\newcommand{\mevc}{\ensuremath{\aunit{Me\kern -0.1em V\!/}c}\xspace}
\newcommand{\gevc}{\ensuremath{\aunit{Ge\kern -0.1em V\!/}c}\xspace}
\newcommand{\mevcc}{\ensuremath{\aunit{Me\kern -0.1em V\!/}c^2}\xspace}
\newcommand{\gevcc}{\ensuremath{\aunit{Ge\kern -0.1em V\!/}c^2}\xspace}

\def\mum  {\ensuremath{\,\upmu\nospaceunit{m}}\xspace}

\def\gsim{{~\raise.15em\hbox{$>$}\kern-.85em
          \lower.35em\hbox{$\sim$}~}\xspace}
\def\lsim{{~\raise.15em\hbox{$<$}\kern-.85em
          \lower.35em\hbox{$\sim$}~}\xspace}

\def\pt         {\ensuremath{p_{\mathrm{T}}}\xspace}

\def\geant      {\mbox{\textsc{Geant4}}\xspace}

\def\pythia     {\mbox{\textsc{Pythia}}\xspace}

\def\tell1  {TELL1\xspace}
\def\ukl1   {UKL1\xspace}

\newcommand{\lhcborcid}[1]{\href{https://orcid.org/#1}{\hspace*{0.1em}\raisebox{-0.45ex}{\includegraphics[width=1em]{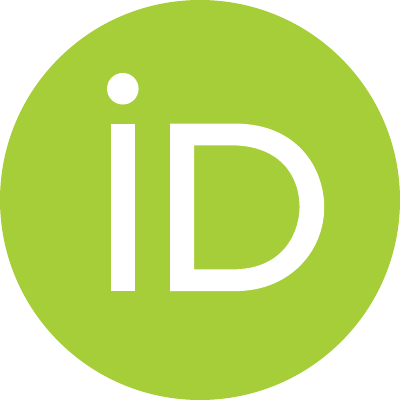}}}}

\usepackage{cite} 
\usepackage{mciteplus}

\usepackage{longtable} 

\begin{document}
\renewcommand{\thefootnote}{\fnsymbol{footnote}}
\setcounter{footnote}{1}


\begin{titlepage}
\pagenumbering{roman}

\vspace*{-1.5cm}
\centerline{\large EUROPEAN ORGANIZATION FOR NUCLEAR RESEARCH (CERN)}
\vspace*{1.5cm}
\noindent
\begin{tabular*}{\linewidth}{lc@{\extracolsep{\fill}}r@{\extracolsep{0pt}}}
\ifthenelse{\boolean{pdflatex}}
{\vspace*{-1.5cm}\mbox{\!\!\!\includegraphics[width=.14\textwidth]{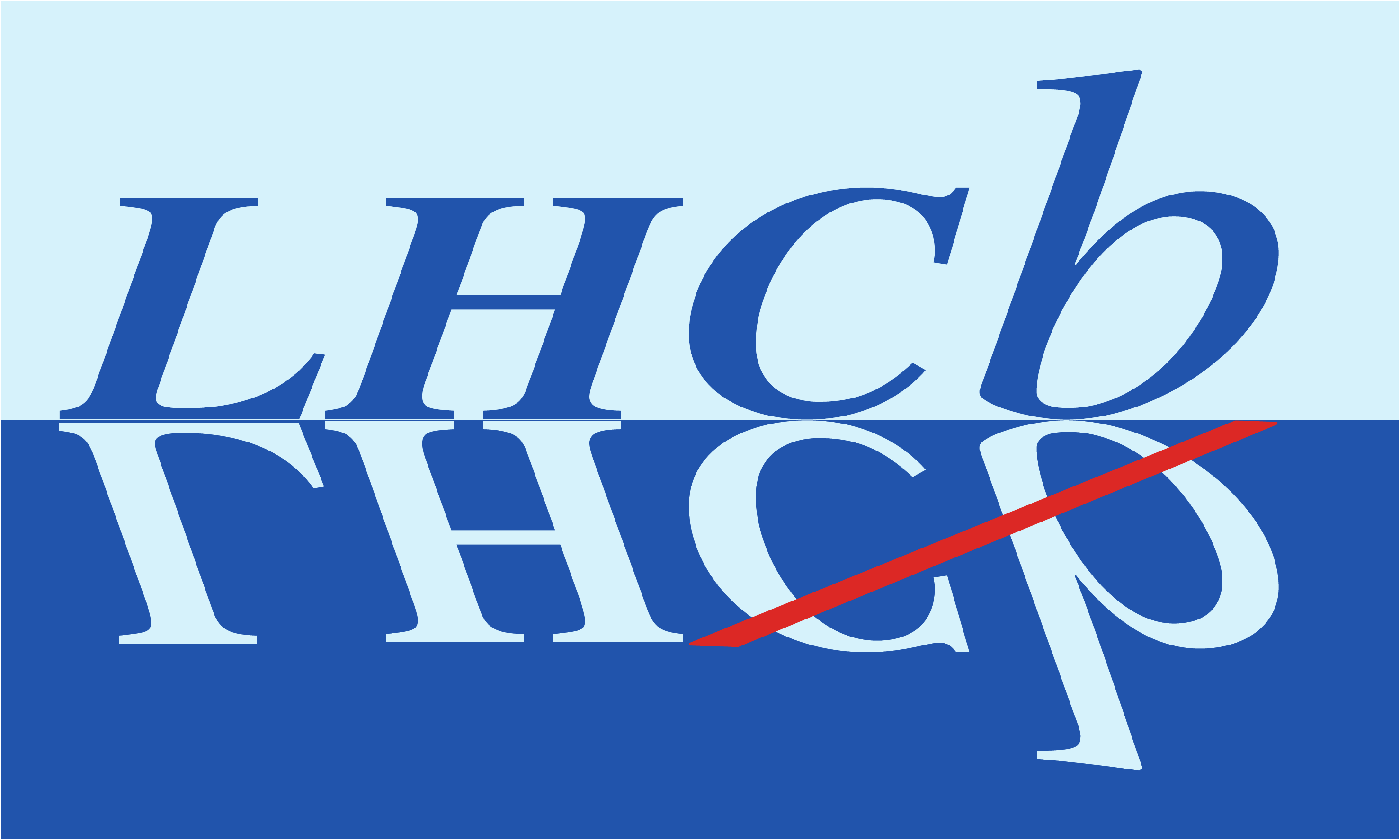}} & &}%
{\vspace*{-1.2cm}\mbox{\!\!\!\includegraphics[width=.12\textwidth]{figs/lhcb-logo.eps}} & &}%
\\
 & & CERN-EP-2023-246 \\  
 & & LHCb-DP-2023-001 \\  
 & & \today \\ 
 & & \\
\end{tabular*}

\vspace*{4.0cm}

{\normalfont\bfseries\boldmath\huge
\begin{center}
  \papertitle 
\end{center}
}

\vspace*{1.5cm}

\begin{center}
\paperauthors\footnote{Authors are listed at the end of this paper.}
\end{center}

\vspace{\fill}

\begin{abstract}
  \noindent
  Momentum measurements for very high momentum charged particles, such as muons from electroweak vector boson decays, are particularly susceptible to charge-dependent curvature biases that arise from misalignments of tracking detectors. Low momentum charged particles used in alignment procedures have limited sensitivity to coherent displacements of such detectors, and therefore are unable to fully constrain these misalignments to the precision necessary for studies of electroweak physics. Additional approaches are therefore required to understand and correct for these effects. In this paper the curvature biases present at the LHCb detector are studied using the pseudomass method in proton-proton collision data recorded at centre of mass energy $\sqrt{s}=13$~TeV during 2016, 2017 and 2018. The biases are determined using $Z\to\mu^+\mu^-$ decays in intervals defined by the data-taking period, magnet polarity and muon direction. Correcting for these biases, which are typically at the $10^{-4}$~GeV$^{-1}$ level, improves the $Z\to\mu^+\mu^-$ mass resolution by roughly 18\% and eliminates several pathological trends in the kinematic-dependence of the mean dimuon invariant mass.
  
\end{abstract}

\vspace*{1cm}

\begin{center}
  Published in 
  JINST 19 (2024) P03010
\end{center}

\vspace{\fill}

{\footnotesize 
\centerline{\copyright~\papercopyright. \href{\paperlicenceurl}{\paperlicence}.}}
\vspace*{2mm}

\end{titlepage}


\newpage
\setcounter{page}{2}
\mbox{~}


\renewcommand{\thefootnote}{\arabic{footnote}}
\setcounter{footnote}{0}

\cleardoublepage

\pagestyle{plain} 
\setcounter{page}{1}
\pagenumbering{arabic}
\section{Introduction}
\newcommand{\nbins}{2048\xspace}
The LHCb experiment~\cite{LHCb-DP-2008-001} is a single-arm dipole spectrometer that is optimised for the study of hadrons containing $b$ or $c$ quarks. Its unique forward-rapidity coverage has also enabled a rich programme studying electroweak and strong processes. A particular challenge in the study of \PW and \PZ boson\footnote{Throughout this paper the term `\PZ boson' includes the effects of the virtual photon, and the interference between the virtual photon and the \PZ boson.} decays to muonic final states is charge-dependent curvature biases in momentum measurements. This paper reports an application of the pseudomass method first described in Ref.~\cite{Pseudomass} to study these biases and to improve the momentum resolution achieved at the LHCb detector for high momentum tracks. This method is complementary to studies of the momentum scale~\cite{LHCb-DP-2023-003}. 
The pseudomass method has already been applied in the first measurement of the $\PW$ boson mass by LHCb~\cite{LHCb-PAPER-2021-024}, which was based on the data recorded during 2016.\footnote{Other electroweak measurements at LHCb have also applied empirical corrections to reduce the impact of these biases.} This study extends upon that analysis by including data recorded in 2017 and 2018, describing the method in greater detail, and showing the performance of the derived corrections with $\PZ\to\mup\mun$ decays. The improvements in mass resolution for $\jpsi\to\mup\mun$ and $\Upsilon(1S) \to \mup\mun$ decays are also evaluated.

Since the start of Run~2 (2015--2018), the LHCb detector has been aligned and calibrated in real time~\cite{LHCb-DP-2019-001,Borghi:2291110}. The tracker alignment is primarily based on charged particles measured in minimum-bias events. A systematic shift of the tracking stations upstream and/or downstream of the magnet is only weakly constrained by this procedure~\cite{Amoraal:2012qn}. Therefore, additional constraints are provided by decays of, for example, \Dz mesons. However, small residual misalignments can still lead to degraded resolution for the very high momentum\footnote{In this publication natural units where $c=1$ are used.} (up to $\mathcal{O}(1~\mathrm{TeV})$) particles produced in the decays of electroweak bosons in the forward region, and can lead to large systematic uncertainties in, for example, the measurement of the $\PW$ boson mass~\cite{LHCb-PAPER-2021-024} or the weak mixing angle~\cite{LHCb-PAPER-2015-039}. 

Crucial for the measurement of charged particle momenta is the magnetic field of the LHCb detector. The magnetic field deflects charged particles, with the curvature of the particles in the field allowing a measurement of $q/p$, where $q$ is the charge of a particle and $p$  its momentum. The LHCb experiment is run in two different configurations: one with the magnetic field pointing upwards, and one with the magnetic field pointing downwards. The first configuration bends positively charged particles
in the horizontal plane towards the centre of the LHC ring, while the second bends negatively charged particles in this direction. The magnetic field used at LHCb is discussed further in Ref.~\cite{LHCb-DP-2008-001}.

The leading biases in the measurement of the momentum follow
\begin{equation}
\label{eq:bias}
    \frac{q}{p} \to \frac{q}{p^\prime} = \frac{q}{\alpha p} + \delta,
\end{equation}
where $p^\prime$ is the `biased' momentum, and $p$ is unbiased. Here $\delta$ is a curvature bias that provides a charge-dependent momentum bias, and $\alpha$ is a factor that sets the overall momentum scale. If these parameters can be accurately determined, then measurements can be corrected for these effects, returning unbiased estimates of particle momenta. This paper considers the corrections necessary to remove the biases associated with $\delta$. The momentum scale factor arises from inaccuracies in the magnetic field calibration and/or knowledge of the length of the detector along the beam ($z$) axis while curvature biases arise from misalignments of the tracking detectors along the bending ($x$) direction.\footnote{The $x$-axis is oriented horizontally towards the outside of the LHC ring.} In general, both parameters can vary across the detector, and with time. The $\delta$ parameter for a given point within the detector is expected to change sign whenever the polarity of the dipole magnetic field at LHCb is reversed, though this reversal can also lead to mechanical displacements of the detector elements.

\begin{figure}[!htbp]
  \begin{center}
      \includegraphics[width=0.8\textwidth]{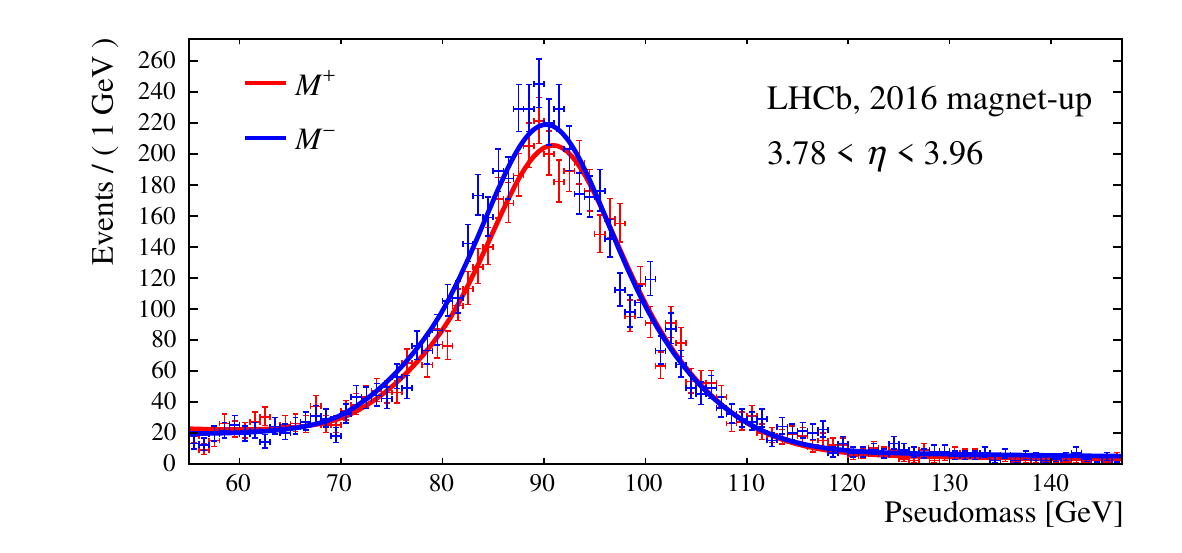}
  \end{center}
  \caption{Pseudomass distributions in the 2016 magnet-up dataset with muons in the pseudorapidity interval $3.78 < \eta< 3.96$. The selection of the data is discussed in Sec.~\ref{sec:selection}, while the function fit to the distributions is defined in Sec.~\ref{sec:fits}. }
  \label{fig:pseudomass}
\end{figure}

The $\alpha$ parameter can be determined from measurements of the masses of known resonances. For example, in decays to a $\mup\mun$ pair, assuming massless muons, the measured mass scales as the square root of the product of the two $\alpha$ values corresponding to the muon trajectories.
On the other hand, in dilepton studies the effects of finite $\delta$ values cancel to a large degree when integrating over the full detector. Despite this calculation, it remains important to determine the $\delta$ values since such biases are important in processes that involve a single lepton (e.g. measurements of the $W$ boson mass). In addition, if uncorrected, the presence of the $\delta$ term also degrades the momentum resolution of the detector.  In Ref.~\cite{Pseudomass} it was proposed to measure the $\delta$ values by defining a so-called pseudomass for resonance decays to $\mup\mun$ pairs. This variable is defined as
\begin{equation}
\mathcal{M}^{\pm} \equiv  \sqrt{\frac{p_{\rm{T}}^{\pm}}{p_{\rm{T}}^{\mp}}}M 
 = \sqrt{2p^{+}p^{-}\frac{p_{\rm{T}}^{\pm}}{p_{\rm{T}}^{\mp}}\left(1 - \cos\theta\right)} 
 = \sqrt{2p^{\pm}p_{\rm{T}}^{\pm}\frac{p^{\mp}}{p_{\rm{T}}^{\mp}}\left(1 - \cos\theta\right)}, 
\end{equation}
where $\theta$ is the opening angle between the two muons, $M$ is the invariant mass of the dimuon system, and \pt is the momentum of a particle transverse to the beam axis. Crucially, since the ratio $\frac{p_{\mp}}{p_{\rm{T}}^{\mp}}$ is completely defined by the measured direction of one muon, $\mathcal{M^\pm}$ solely depends on the momentum measurement of the other muon, and therefore only on the value of $\delta$ associated with the regions of the detector that the corresponding muon traverses. This allows the effect of local detector misalignments to be considered. Figure~\ref{fig:pseudomass} shows example $\mathcal{M^\pm}$ distributions in the LHCb Run~2 dataset, with the relevant positively or negatively charged muon in the pseudorapidity\footnote{The pseudorapidity of a particle is defined as $\eta\equiv -\ln(\tan(\frac{\vartheta}{2}))$, where $\vartheta$ is the polar angle of the particle with respect to the colliding beam direction.}  interval $3.78 < \eta< 3.96$. 

The $\PZ\to\mup\mun$ cross-section is largest where the transverse momentum of the dimuon system is much smaller than the mass. In this low transverse momentum limit, where $p_{\rm{T}}^{+} \sim p_{\rm{T}}^{-}$, $\mathcal{M^+}$ and $\mathcal{M^-}$ approximate $M$ such that the distributions shown in Fig.~\ref{fig:pseudomass} peak at the \PZ boson mass. By calculating the pseudomass in the presence of curvature biases of the form given in Eq.~\ref{eq:bias}, it can be seen that the presence of such biases can be directly related to an asymmetry $\mathcal{A}$ between the peak positions in the $\mathcal{M^+}$ and  $\mathcal{M^-}$ distributions through
\begin{equation}
\label{eq:asymmetry}
\delta \approx -\frac{\mathcal{A}}{2}\times \left( \Bigl\langle\frac{1}{p^{+}}\Bigr\rangle+\Bigl\langle\frac{1}{p^{-}}\Bigr\rangle \right), 
\end{equation}
where $\Bigl\langle\frac{1}{p^{\pm}}\Bigr\rangle$ are the mean values of the inverse of the muon momenta. The asymmetry in peak positions is evaluated as $\mathcal{A}=(u^+ - u^-)/(u^+ + u^-)$, where $u^\pm$ is the mean parameter associated with a peaking component in a fit to the $\mathcal{M}^\pm$ distributions. Section~\ref{sec:fits} discusses the functional form used to describe the $\mathcal{M}^\pm$ distributions in more detail. The evaluation of this asymmetry therefore allows the value of $\delta$ to be found using Eq.~\ref{eq:asymmetry}, which in turn allows curvature bias corrections to be derived and applied by inverting Eq.~\ref{eq:bias}. The key feature of this method is that these corrections can be independently determined in arbitrarily small regions of the detector, since $\mathcal{M^\pm}$ only depends on the momentum of one of the two muons. The approximation that $p_{\rm{T}}^{+} \sim p_{\rm{T}}^{-}$ can be improved, for example by rejecting candidates with large values of the angular variable $\phi^{\ast}$~\cite{phistar}, defined as:

\begin{equation}
\label{eq:curvatureBias}
\phi^{\ast}  \equiv  \tan \left(\frac{\pi-\Delta\phi}{2}\right) \sin\theta^{*}_{\eta}, 
\end{equation}
where $\Delta\phi$ is the azimuthal opening angle between the two leptons in the transverse plane. The quantity $\theta^*_\eta$ is defined by $\cos \theta^{*}_{\eta} \equiv \tanh{\frac{\eta^{-} - \eta^{+}}{2}}$, where $\eta^{-}$ and $\eta^{+}$ are the pseudorapidities of the negatively and positively charged leptons, respectively.

\section{Dataset, simulation samples and signal selection}
\label{sec:selection}
This paper considers $pp$ collision data recorded during 2016--2018 at $\sqrt{s}=13$~TeV, corresponding to an integrated luminosity of 5.3\,fb$^{-1}$. The detector includes a high-precision tracking system consisting of a silicon-strip vertex detector surrounding the $pp$ interaction region~\cite{LHCb-DP-2014-001}, a large-area silicon-strip detector located upstream of a dipole magnet with a bending power of about $4{\mathrm{\,Tm}}$, and three stations of silicon-strip detectors and straw drift tubes~\cite{LHCb-DP-2017-001} placed downstream of the magnet. 
Muons are identified by a system composed of alternating layers of iron and multi-wire proportional chambers~\cite{LHCb-DP-2012-002}. The online event selection is performed by a trigger~\cite{LHCb-DP-2012-004}, which consists of a hardware stage based on information from the calorimeter and muon systems, followed by two software stages which use a full event reconstruction~\cite{LHCb-DP-2019-001}.

In the simulation, $pp$ collisions are generated using \pythia~\cite{Sjostrand:2007gs,Sjostrand:2006za} with a specific \lhcb configuration~\cite{LHCb-PROC-2010-056}.
The interaction of the generated particles with the detector, and its response, are implemented using the \geant toolkit~\cite{Allison:2006ve, *Agostinelli:2002hh} as described in Ref.~\cite{LHCb-PROC-2011-006}. 

The selection of candidate $\PZ\to\mup\mun$ decays follows that described in detail in Ref.~\cite{LHCb-PAPER-2021-024}. Tracks are reconstructed in the LHCb detector following Ref.~\cite{LHCb-DP-2013-002}. Candidate decays are formed from oppositely charged pairs of muons, each with $\pt > 20$\gev, and in the range $1.9<\eta<4.9$, and where each pair forms a common decay vertex. Loose requirements on the track fit $\chi^2$, impact parameter and track isolation suppress background from misidentified hadrons and heavy-flavour hadron decays. The dimuon candidate is required to have an invariant mass above 40\gev, though additional requirements are detailed further in Sec.~\ref{sec:fits} when applying the pseudomass method. At least one of the two muons must be matched to a positive decision of a single muon selection in all three stages of the trigger. When deriving the pseudomass corrections, candidate events are required to have $\phi^{\ast} < 0.1$. The purity of the $\PZ\to\mup\mun$ sample is studied following Ref.~\cite{LHCb-PAPER-2021-024}, and background contributions are found to be at the per-mille level.

Candidate $\jpsi\to\mup\mun$ and $\Upsilon(1S) \to \mup\mun$ decays are also used to evaluate the impact of the curvature-bias corrections on mass resolutions. The selection of these candidates again follows Ref.~\cite{LHCb-PAPER-2021-024}. These candidates are formed from oppositely charged pairs of muons, satisfying basic quality requirements that form a common decay vertex. The muons from $\jpsi$ and $\Upsilon(1S)$ decays are required to have \pt values larger than 0.5\gev and 3\gev, respectively. The $\jpsi$ decay vertices are required to be significantly displaced from their associated primary interaction vertex, such that the signal is almost entirely produced from decays of $b$ hadrons.
The $\jpsi$ and $\Upsilon(1S)$ candidates must be matched to positive decisions of muon selections in all three stages of the trigger.

\section{Application of the pseudomass method}
\label{sec:fits}
The data are analysed independently for each of the three years studied in LHC Run~2 (2016--2018) and for each of the two magnet polarity configurations in each year.
The 2018 dataset is further divided into two periods, since the real-time alignment procedures were re-optimised in the later part of that year.
The data are further divided into 256 directional intervals, corresponding to the product of 16 intervals in the range $1.9 < \eta < 4.9$ and 16 intervals in the range $-\pi < \phi < \pi$. In total, there are \nbins intervals in which the curvature biases are independently evaluated. 

Histograms of the pseudomass distributions, covering the range $55 < \mathcal{M}^{\pm} < 147$\gev, are produced for each of the above intervals. 
For each of these intervals, the mass asymmetry parameters $\mathcal{A}$ are determined with a simultaneous maximum-likelihood fit to the two distributions. 
The data are sufficiently pure such that they can be modelled considering signal alone without modelling any background contribution.
The model for each pseudomass distribution includes a non-resonant virtual photon component described by an exponential function, and a resonant component described by the sum of two Crystal Ball functions, with shared means, but with differing widths. Both the fraction of the non-resonant and one of the Crystal Ball components vary freely in the fit.  Figure~\ref{fig:pseudomass} shows the pseudomass distributions for the $3.78 < \eta < 3.96$ interval, combining all $\phi$ intervals, with the fit results overlaid. The means of the Crystal Ball functions describing the $\mathcal{M}^{\pm}$ distributions are parameterised as a charge-averaged mass, which varies freely in the fits, multiplied by $1 \pm \mathcal{A}$, where $\mathcal{A}$ also varies freely.

\begin{figure}[!htbp]
  \begin{center}
      \includegraphics[width=0.99\textwidth]{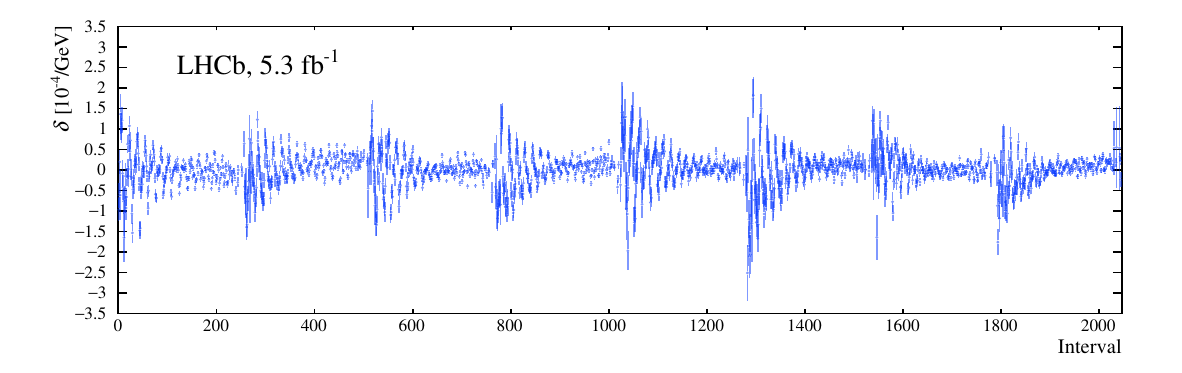}
      \includegraphics[width=0.99\textwidth]{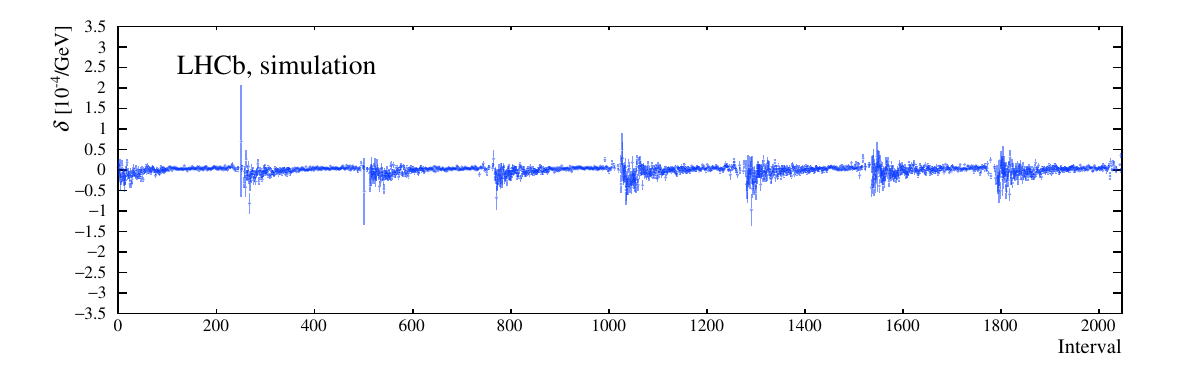}
  \end{center}
    \caption{Curvature biases for LHCb (upper) data and (lower) simulation. The \nbins intervals are numbered as $i_\phi + (i_\eta \times 16) + ( i_\text{polarity} \times 16 \times 16) + ( i_\text{period} \times 16 \times 16 \times 2 )$, where $i_\phi$, $i_\eta$, $i_\text{polarity}$ and $i_\text{period}$ individually index the azimuthal angle $\phi$, pseudorapidity, polarity of the magnetic field, and the data-taking period. The uncertainties shown are statistical.}
  \label{fig:correction_index}
\end{figure}

The resulting asymmetry values are translated to curvature biases according to Eq.~\ref{eq:asymmetry}. These biases are presented in Fig.~\ref{fig:correction_index}, for both data and simulation. The periodic repeating pattern is caused by the measured biases being similar in different data taking periods. Since the furthest forward region typically contains higher momentum tracks, small biases are better controlled by the standard LHCb alignment techniques. Consequently, the residual bias determined in data using the pseudomass approach typically decreases as the pseudorapidity increases.

\section{Biases from fundamental physics}
\label{sec:physics}
In the $\PZ\to\mup\mun$ process the two muons are not completely equivalent; the $p_{\rm{T}}^{+} \sim p_{\rm{T}}^{-}$ approximation is not perfect. There is a `forward-backward' asymmetry introduced by the interference of the \PZ boson and the virtual photon, as well as from the vector and axial-vector couplings of the \PZ boson. 
While the primary impact of this physical effect is to change the yields in each muon pseudorapidity interval, it does also lead to small natural asymmetries in the peak positions of the $\mathcal{M}^+$ and $\mathcal{M}^-$ distributions that should not be corrected for.
In the simulation, in which no misalignment is present, the extracted curvature bias stems from this physical effect. This is confirmed by explicitly examining the $q/p$ residuals with respect to~the generated values, with and without curvature bias corrections applied. Therefore, it can be concluded from Fig.~\ref{fig:correction_index} that the effect of the physics bias ($\mathcal{O}(10^{-5} - 10^{-6}$\,GeV$^{-1})$ is much smaller than the measured curvature biases in the data. In order to remove the effects of the physics bias, the bias observed in simulation is subtracted from the one measured in data.

\begin{figure}[!htbp]
  \begin{center}

      \includegraphics[width=0.9\textwidth]{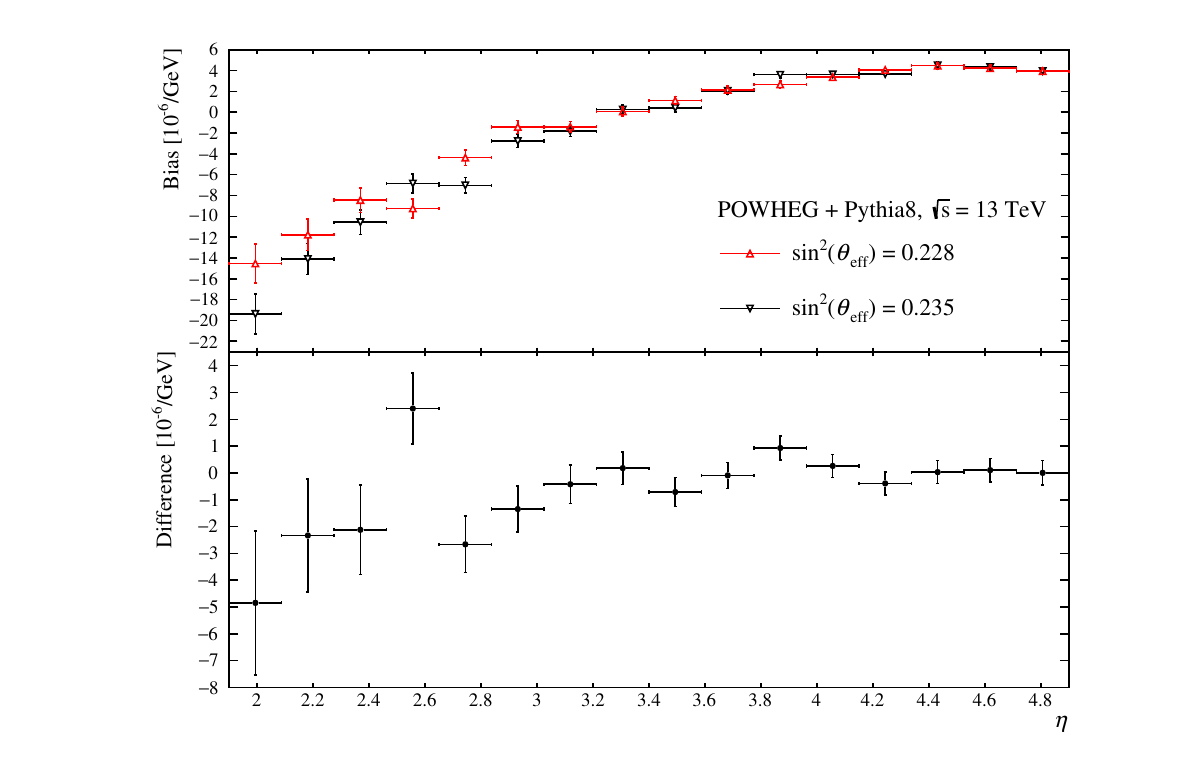}
  \end{center}
    \caption{Upper: pseudomass corrections derived from simulation for two different values of $\sin^2(\theta_{\rm eff})$. Since no misalignment is present in these studies this shows a fundamental physics bias associated with the method. Lower: the difference between the biases found in the two samples. The uncertainties shown are statistical.}
  \label{fig:physicsbias}
\end{figure}

The extracted physics bias is influenced by the value of the weak mixing angle used in the simulation. This was already studied in Ref.~\cite{Pseudomass}, where effects were shown to be small. A more detailed investigation is considered here. The pseudomass method is applied directly to $\PZ\to\mup\mun$ events simulated using \textsc{Powheg-Box}~\cite{POWHEGewZ,powheg1,powheg2,powheg3} interfaced with \textsc{Pythia8}~\cite{Sjostrand:2007gs,Sjostrand:2006za,Pythia8Main}. The simulated events have no detector simulation applied but the muon momenta are smeared according to previously measured resolutions~\cite{LHCb-DP-2014-002}. Curvature biases are determined for two different values of the weak mixing angle $\sin^2(\theta_\text{eff}) = 0.228$ and $0.235$; the difference between these values is approximately 40 times larger than the uncertainty in the world average~\cite{ALEPH:2005ab}, or about $\pm 9$ times the precision of the most precise hadron collider measurements of this parameter to date. The use of a wide spacing between the two samples allows more statistical power in testing for the dependence of the physics bias on the input values of the weak mixing angle. Figure~\ref{fig:physicsbias} shows that the curvature biases are broadly consistent with those found using the full LHCb simulation, much smaller than the corrections derived in the data, and are consistent between the two values of the weak mixing angle considered. The modelling of this physics effect is therefore found to have a negligible impact on the understanding of the detector alignment.

\section{Results}
\label{sec:results}
\begin{figure}[!tbp]
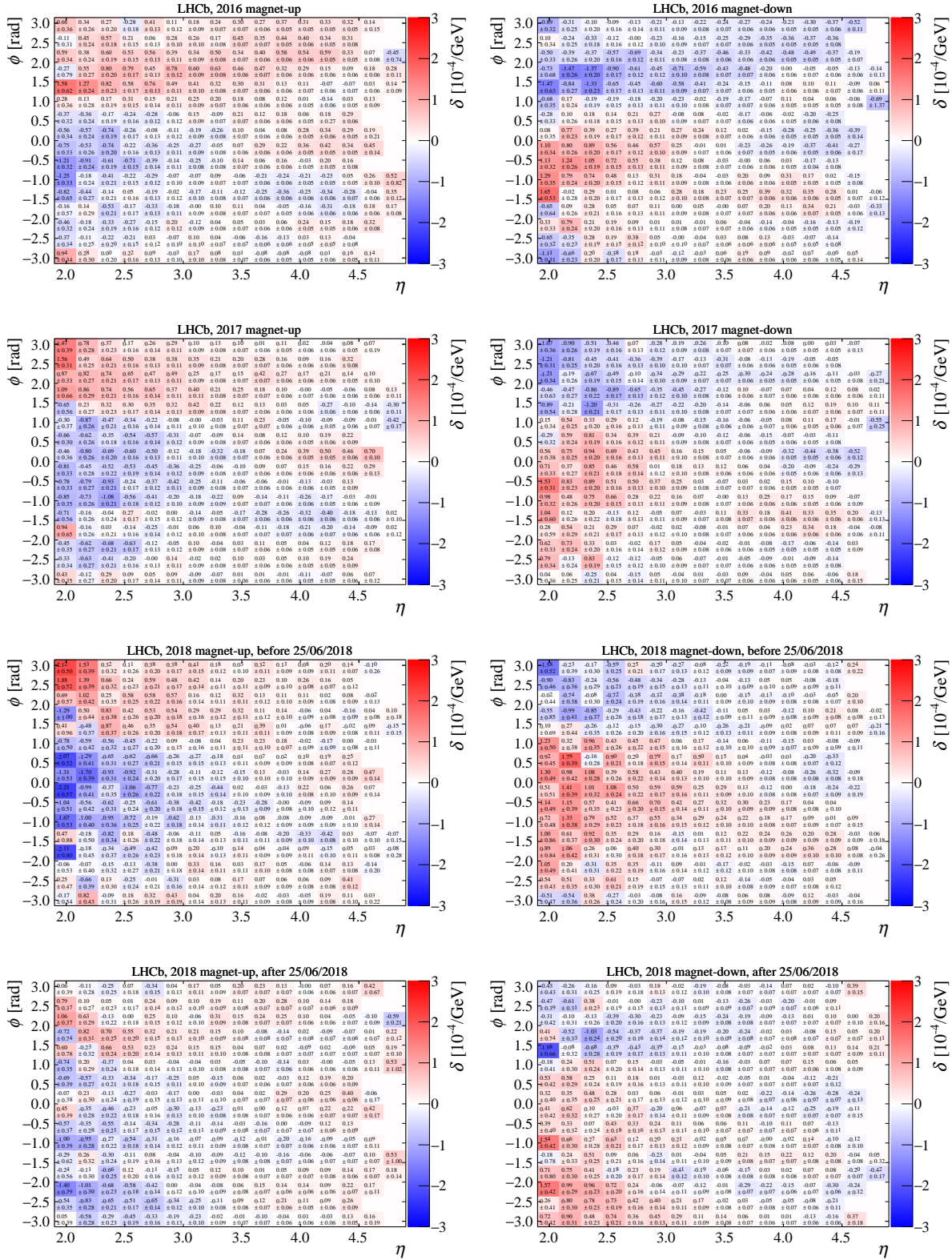

  \begin{center}
  \foreach \period in {0,1,2,3}{
    \includegraphics[width=\figwidth]{plots/pseudomass_corrections-map__d13TeV__DATA_Minus_MC__\period__up__COLZ_TEXTE.pdf}
    \includegraphics[width=\figwidth]{plots/pseudomass_corrections-map__d13TeV__DATA_Minus_MC__\period__down__COLZ_TEXTE.pdf}
  }
  \end{center}
  \caption{Curvature biases as a function of $\eta$ and $\phi$. The biases are shown as a `heat map', with the colour-scale showing the size of the biases. The left and right columns correspond to the up- and down-polarity configurations, respectively, while the rows correspond to the 2016, 2017, 2018 (two periods) datasets. Empty regions at larger $\eta$ values are due to the rectangular acceptance of the detector.}
\label{fig:correction_map_data_minus_mc}
\end{figure}

The curvature biases are shown as a function of $\eta$ and $\phi$, separately for the two magnet polarities and four data-taking periods, in Fig.~\ref{fig:correction_map_data_minus_mc}, after subtracting the physics biases determined in simulation, as discussed in Sec.~\ref{sec:physics}.

\begin{figure}[!htbp]
  \begin{center}
    \includegraphics[width=\figwidth]{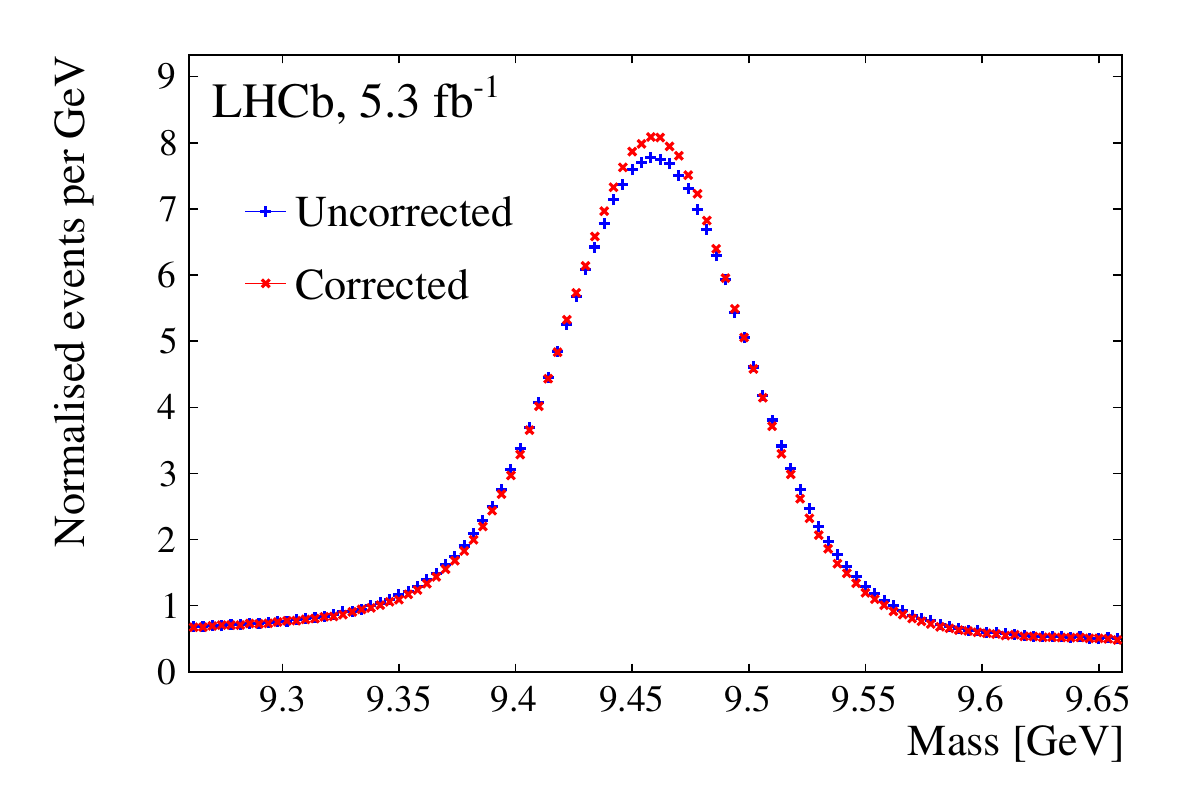}
    \includegraphics[width=\figwidth]{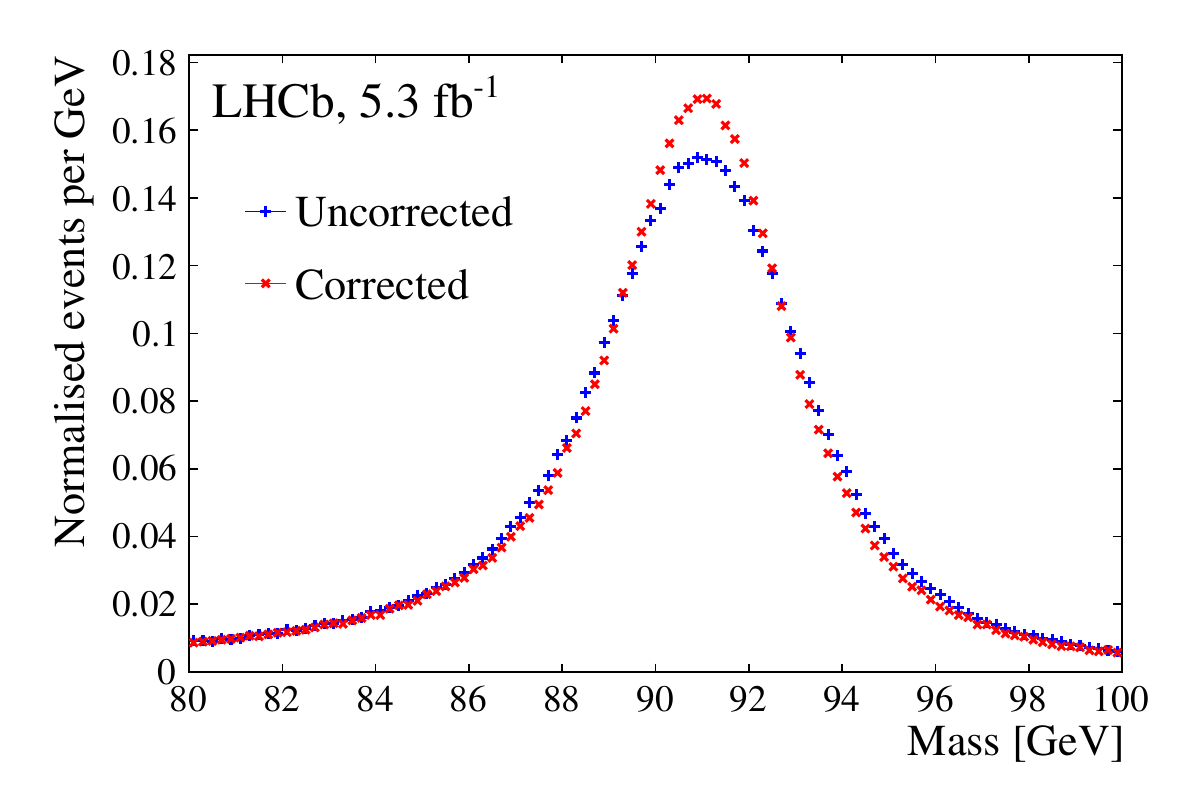}
  \end{center}
    \caption{Mass distributions for the (left) $\Upsilon(1S)$ and (right) $\PZ$ resonances. The data are shown with and without the corrections applied.}
  \label{fig:correction_on_mass}
\end{figure}

\begin{figure}[!htbp]
  \begin{center}
      \includegraphics[width=0.8\textwidth]{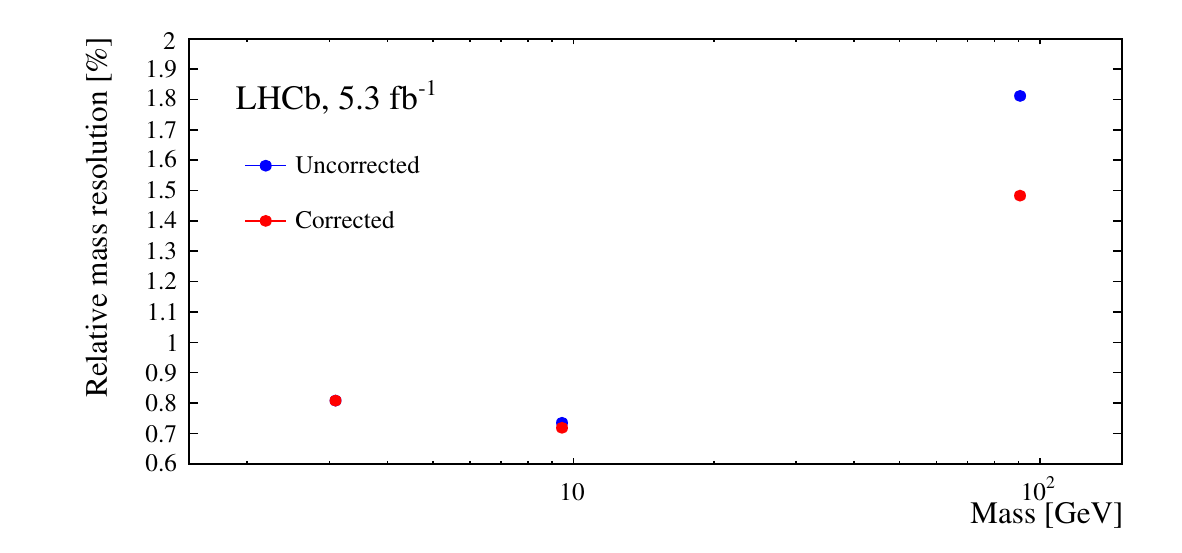}
  \end{center}
    \caption{Relative mass resolution as a function of the dimuon invariant mass, showing the resolution of the $\jpsi$, $\Upsilon(1S)$ and $\PZ$ resonances with and without the corrections applied. Statistical uncertainties on these resolution values are too small to be visible in the plot. }
  \label{fig:correction_resolution}
\end{figure}

Figure~\ref{fig:correction_on_mass} shows the mass distributions of the $\Upsilon(1S) \to \mup\mun$ and $\PZ \to \mup\mun$ candidates, with and without the application of the curvature bias corrections, while Fig.~\ref{fig:correction_resolution} shows estimates of the relative mass resolutions for the three resonances before and after the corrections are applied. All years and both magnet polarities are combined. The $\PZ \to \mup\mun$ mass resolution is estimated with a parametric fit of the sum of an exponential function and a relativistic Breit-Wigner function with the $\PZ$ boson resonance parameters, convolved with a Gaussian resolution function. The resolution of the $\PZ \to \mup\mun$ mass peak is improved by 18\% using the method presented here. 
The $\Upsilon(1S)$ resonance is sufficiently narrow such that the observed width can be assumed to be the detector resolution. The $\Upsilon(1S) \to \mup\mun$ mass resolution is improved using the method presented here by 2.3\%.
There is no significant effect on the resolution of the $\jpsi \to \mup\mun$ mass, since the correspondingly low momentum muons are less susceptible to curvature biases. The improvements in resolution are found to be robust when the data are split randomly, with independent datasets used to find the corrections and then to test their impact.

Systematic misalignments of the detector also lead to non-physical trends in the mean dimuon mass of $\PZ$ boson events as a function of angular variables perpendicular to the beamline. In Fig.~\ref{fig:profileplots_mum-phi} it can be seen that this mass varies strongly with the azimuthal angle of the negatively charged muon. This trend roughly changes in sign with the magnet polarity. The mean dimuon invariant mass varies within $\pm 2$\gev in the 2016 and 2017 datasets, while the variations are reduced by roughly a factor of two in the 2018 data. The reduction in variation in the 2018 dataset is explained by the $\PZ$ boson data being used in the real-time alignment in the latter part of this period. However, even with the use of this data in real-time the pseudomass method still provides significant improvements in the observed trends. These trends are consistent with a relative shift along the $x$ axis of the upper and lower sections of the tracking detectors. A similar trend is visible in other related variables such as the $\phi$ of the positively charged muon. The curvature-bias corrections presented here eliminate these trends, and no such trend is visible after the corrections are applied.

\begin{figure}[!htbp]
  \begin{center}
  \foreach \dataset in {2016-up,2016-down,2017-up,2017-down,2018-up,2018-down}{    \includegraphics[width=\figwidth]{plots/pseudomass_corrections-profile__d13TeV__\dataset__mum-phi}
   }
   \end{center}
   \caption{Mean dimuon mass as a function of the azimuthal angle of the negatively charged muon for $\PZ \to\mup\mun$ candidates. The left and right hand columns correspond to the up and down polarities of the magnetic field while the rows correspond to the 2016, 2017 and 2018 datasets.}
\label{fig:profileplots_mum-phi}
\end{figure}

\begin{figure}[!htbp]
  \begin{center}
  \foreach \dataset in {2016-up,2016-down,2017-up,2017-down,2018-up,2018-down}{    
  \includegraphics[width=\figwidth]{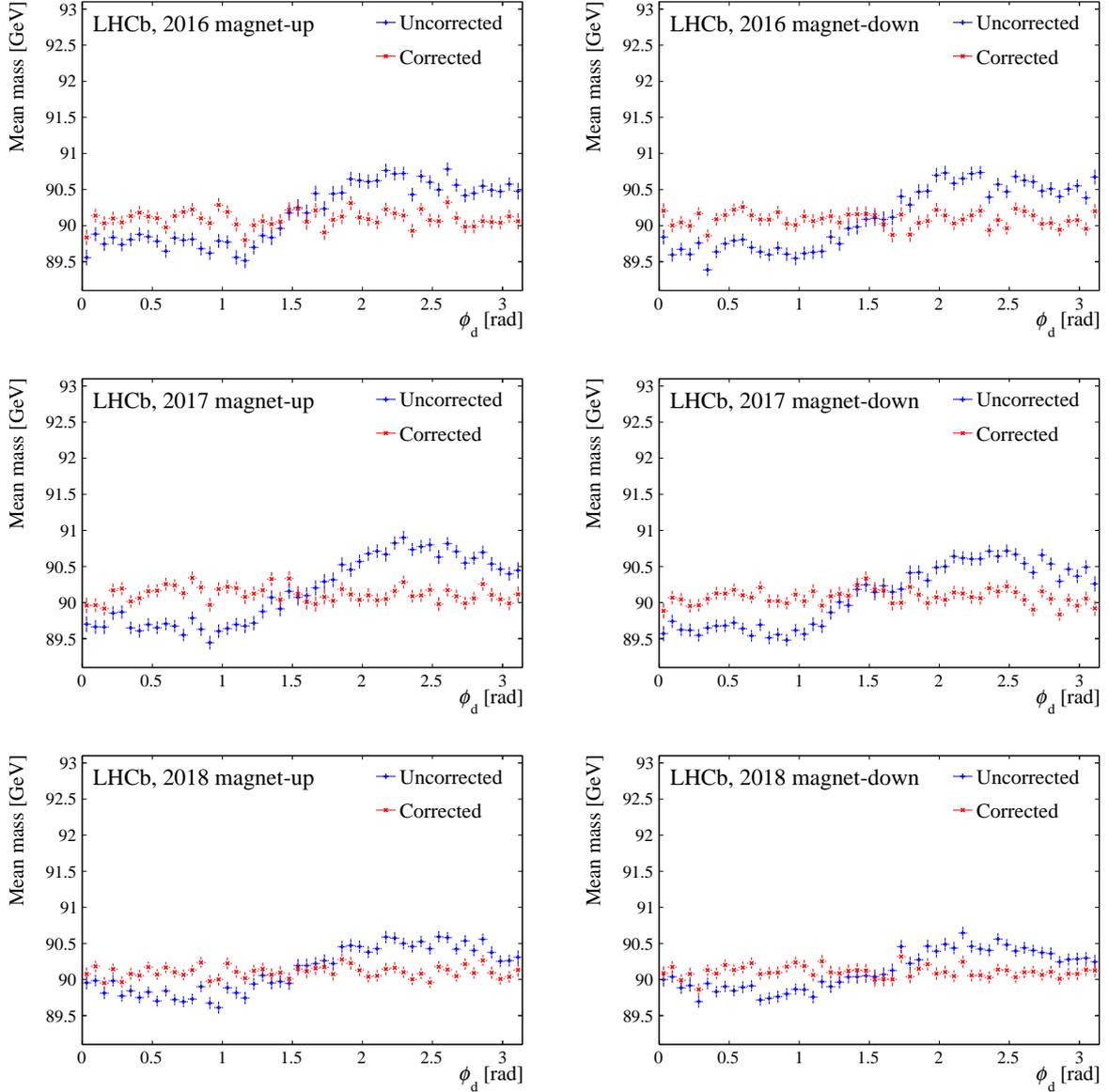}
  }
   \end{center}
   \caption{Mean dimuon mass as a function of $\phi_{\rm d}$ for $Z \to\mup\mun$ candidates. The left and right-hand columns correspond to the up- and down-polarities of the magnetic field while the rows correspond to the 2016, 2017 and 2018 datasets.}
\label{fig:profileplots_mum-phid}
\end{figure}

A particularly interesting variable in the study of curvature biases is the angle
\begin{equation}
    \phi_{\rm d} = \cos^{-1}\left( \frac{ \vec{D} \cdot \hat{\vec{B}}}{|\vec{D}|}\right),
\end{equation}
where $\vec{D}$ is a decay plane vector defined from the three-momenta of the $\mup$ and $\mun$ as
\begin{equation}
\vec{D} = \vec{p}_{+} \times \vec{p}_{-},
\end{equation}
and $\hat{\vec{B}}$ is a magnetic field three-vector defined as $(0,1,0)$ and $(0,-1,0)$ for the up- and down-polarity configurations, respectively. This variable measures the angle between the normal to the dimuon decay plane and the magnetic field. Configurations in which the two muons bend towards and away from each other in the detector also take different values.
Figure~\ref{fig:profileplots_mum-phid} shows a variation in the mean dimuon mass in $\PZ\to\mup\mun$ candidates of around $1$\gev before the corrections discussed here are applied. This is characteristic of a relative displacement along the beam axis of the tracking detectors before and after the magnet.  The method presented here recovers a profile of the mean dimuon mass that is independent of $\phi_{\rm d}$.

Simplified fits are performed to investigate the extent to which the pseudomass corrections can be attributed to coherent rotations and shifts of the tracking detectors. This uses the same simplified model of the LHCb detector as Ref.~\cite{Pseudomass}. The fits favour shifts along the $x$ axis of $\mathcal{O}(10 \mum)$, rotations around the $z$ axis of $\mathcal{O}(100~\mathrm{\upmu rad})$ and shifts along the $z$ axis of $\mathcal{O}(0.5~\mathrm{mm})$. These are found to be compatible with the changes applied in the alignment for the latter part of the 2018 dataset. When this simple model is fit to the values of $\delta$ measured and reported in Fig.~\ref{fig:correction_map_data_minus_mc}, allowing for these three transformations simultaneously typically improves the $\chi^2$ that describes the fit by about 1000 units (for 238 degrees of freedom), depending on the year and polarity. However, despite this improvement, the  $\chi^2$ probabilities for the simplified model remain poor, implying that the curvature biases present correspond to a mixture of coherent and local effects. This reaffirms the importance of a model-independent method to determine local curvature biases, such as the pseudomass method presented here.

\section{Conclusions}

The measurement of charged particle momenta by the LHCb experiment is subject to both charge-independent and charge-dependent biases, which must be calibrated. The latter is the focus of this work, and arises due to systematic misalignments of the tracking detectors that are not fully resolved by the real-time alignment procedures. Studies of muonic decays of $\PW$ and $\PZ$
 bosons are particularly susceptible to this effect, since the muons have momenta up to $\mathcal{O}$(TeV). The pseudomass of a two-body decay of a neutral resonance only depends on the momentum of one final state particle, and therefore has first-order sensitivity to the curvature biases. These biases are determined in simultaneous fits of the distributions of the pseudomasses computed from the $\mup$ and $\mun$ momenta. The curvature biases from the experimental data recorded during the 2016, 2017 and 2018 $pp$ collisions are extracted in kinematic intervals for each period. The resulting curvature biases are typically at the level of $\mathcal{O}(10^{-4})$\,GeV$^{-1}$. After applying these corrections to the data, an improvement of the $\PZ\to\mup\mun$ mass resolution of around 18\% is achieved, and the $\Upsilon(1S)\to\mup\mun$ mass resolution is improved by about two percent. 
 In conclusion, the curvature bias corrections described here are essential in measurements of the $\PW$ boson mass and the weak mixing angle with the Run~2 data collected at LHCb, but can also easily be included in analyses of lower mass states.

\section*{Acknowledgements}
%
%
\noindent We express our gratitude to our colleagues in the CERN
accelerator departments for the excellent performance of the LHC. We
thank the technical and administrative staff at the LHCb
institutes.
We acknowledge support from CERN and from the national agencies:
CAPES, CNPq, FAPERJ and FINEP (Brazil); 
MOST and NSFC (China); 
CNRS/IN2P3 (France); 
BMBF, DFG and MPG (Germany); 
INFN (Italy); 
NWO (Netherlands); 
MNiSW and NCN (Poland); 
MCID/IFA (Romania); 
MICINN (Spain); 
SNSF and SER (Switzerland); 
NASU (Ukraine); 
STFC (United Kingdom); 
DOE NP and NSF (USA).
We acknowledge the computing resources that are provided by CERN, IN2P3
(France), KIT and DESY (Germany), INFN (Italy), SURF (Netherlands),
PIC (Spain), GridPP (United Kingdom), 
CSCS (Switzerland), IFIN-HH (Romania), CBPF (Brazil),
and Polish WLCG (Poland).
We are indebted to the communities behind the multiple open-source
software packages on which we depend.
Individual groups or members have received support from
ARC and ARDC (Australia);
Key Research Program of Frontier Sciences of CAS, CAS PIFI, CAS CCEPP, 
Fundamental Research Funds for the Central Universities, 
and Sci. \& Tech. Program of Guangzhou (China);
Minciencias (Colombia);
EPLANET, Marie Sk\l{}odowska-Curie Actions, ERC and NextGenerationEU (European Union);
A*MIDEX, ANR, IPhU and Labex P2IO, and R\'{e}gion Auvergne-Rh\^{o}ne-Alpes (France);
AvH Foundation (Germany);
ICSC (Italy); 
GVA, XuntaGal, GENCAT, Inditex, InTalent and Prog.~Atracci\'on Talento, CM (Spain);
SRC (Sweden);
the Leverhulme Trust, the Royal Society
 and UKRI (United Kingdom).

\addcontentsline{toc}{section}{References}
\bibliographystyle{LHCb}
\bibliography{main,LHCb-PAPER,LHCb-CONF,LHCb-DP,LHCb-TDR,standard}

\ifx\mcitethebibliography\mciteundefinedmacro
\PackageError{LHCb.bst}{mciteplus.sty has not been loaded}
{This bibstyle requires the use of the mciteplus package.}\fi
\providecommand{\href}[2]{#2}
\begin{mcitethebibliography}{10}
\mciteSetBstSublistMode{n}
\mciteSetBstMaxWidthForm{subitem}{\alph{mcitesubitemcount})}
\mciteSetBstSublistLabelBeginEnd{\mcitemaxwidthsubitemform\space}
{\relax}{\relax}

\bibitem{LHCb-DP-2008-001}
LHCb collaboration, A.~A. Alves~Jr.\ {\em et~al.},
  \ifthenelse{\boolean{articletitles}}{\emph{{The \lhcb detector at the LHC}},
  }{}\href{https://doi.org/10.1088/1748-0221/3/08/S08005}{JINST \textbf{3}
  (2008) S08005}\relax
\mciteBstWouldAddEndPuncttrue
\mciteSetBstMidEndSepPunct{\mcitedefaultmidpunct}
{\mcitedefaultendpunct}{\mcitedefaultseppunct}\relax
\EndOfBibitem
\bibitem{Pseudomass}
W.~Barter, M.~Pili, and M.~Vesterinen,
  \ifthenelse{\boolean{articletitles}}{\emph{{A simple method to determine
  charge-dependent curvature biases in track reconstruction in hadron collider
  experiments}},
  }{}\href{https://doi.org/10.1140/epjc/s10052-021-09016-9}{Eur.\ Phys.\ J.\
  \textbf{C81} (2021) 251},
  \href{http://arxiv.org/abs/2101.05675}{{\normalfont\ttfamily
  arXiv:2101.05675}}\relax
\mciteBstWouldAddEndPuncttrue
\mciteSetBstMidEndSepPunct{\mcitedefaultmidpunct}
{\mcitedefaultendpunct}{\mcitedefaultseppunct}\relax
\EndOfBibitem
\bibitem{LHCb-DP-2023-003}
LHCb collaboration, R.~Aaij {\em et~al.},
  \ifthenelse{\boolean{articletitles}}{\emph{{Momentum scale calibration of the
  LHCb spectrometer}}, }{} LHCb-DP-2023-003, {in preparation}\relax
\mciteBstWouldAddEndPuncttrue
\mciteSetBstMidEndSepPunct{\mcitedefaultmidpunct}
{\mcitedefaultendpunct}{\mcitedefaultseppunct}\relax
\EndOfBibitem
\bibitem{LHCb-PAPER-2021-024}
LHCb collaboration, R.~Aaij {\em et~al.},
  \ifthenelse{\boolean{articletitles}}{\emph{{Measurement of the $W$ boson
  mass}}, }{}\href{https://doi.org/10.1007/JHEP01(2022)036}{JHEP \textbf{01}
  (2022) 036}, \href{http://arxiv.org/abs/2109.01113}{{\normalfont\ttfamily
  arXiv:2109.01113}}\relax
\mciteBstWouldAddEndPuncttrue
\mciteSetBstMidEndSepPunct{\mcitedefaultmidpunct}
{\mcitedefaultendpunct}{\mcitedefaultseppunct}\relax
\EndOfBibitem
\bibitem{LHCb-DP-2019-001}
R.~Aaij {\em et~al.}, \ifthenelse{\boolean{articletitles}}{\emph{{Design and
  performance of the LHCb trigger and full real-time reconstruction in Run 2 of
  the LHC}}, }{}\href{https://doi.org/10.1088/1748-0221/14/04/P04013}{JINST
  \textbf{14} (2019) P04013},
  \href{http://arxiv.org/abs/1812.10790}{{\normalfont\ttfamily
  arXiv:1812.10790}}\relax
\mciteBstWouldAddEndPuncttrue
\mciteSetBstMidEndSepPunct{\mcitedefaultmidpunct}
{\mcitedefaultendpunct}{\mcitedefaultseppunct}\relax
\EndOfBibitem
\bibitem{Borghi:2291110}
S.~Borghi, \ifthenelse{\boolean{articletitles}}{\emph{{Novel real-time
  alignment and calibration of the LHCb detector and its performance}},
  }{}\href{https://doi.org/10.1016/j.nima.2016.06.050}{Nucl.\ Instrum.\ Meth.\
  \textbf{A845} (2017) 560}\relax
\mciteBstWouldAddEndPuncttrue
\mciteSetBstMidEndSepPunct{\mcitedefaultmidpunct}
{\mcitedefaultendpunct}{\mcitedefaultseppunct}\relax
\EndOfBibitem
\bibitem{Amoraal:2012qn}
J.~Amoraal {\em et~al.},
  \ifthenelse{\boolean{articletitles}}{\emph{{Application of vertex and mass
  constraints in track-based alignment}},
  }{}\href{https://doi.org/10.1016/j.nima.2012.11.192}{Nucl.\ Instrum.\ Meth.\
  \textbf{A712} (2013) 48},
  \href{http://arxiv.org/abs/1207.4756}{{\normalfont\ttfamily
  arXiv:1207.4756}}\relax
\mciteBstWouldAddEndPuncttrue
\mciteSetBstMidEndSepPunct{\mcitedefaultmidpunct}
{\mcitedefaultendpunct}{\mcitedefaultseppunct}\relax
\EndOfBibitem
\bibitem{LHCb-PAPER-2015-039}
LHCb collaboration, R.~Aaij {\em et~al.},
  \ifthenelse{\boolean{articletitles}}{\emph{{Measurement of the
  forward-backward asymmetry in \mbox{\decay{Z/\gamma^\ast}{\mumu}} decays and
  determination of the effective weak mixing angle}},
  }{}\href{https://doi.org/10.1007/JHEP11(2015)190}{JHEP \textbf{11} (2015)
  190}, \href{http://arxiv.org/abs/1509.07645}{{\normalfont\ttfamily
  arXiv:1509.07645}}\relax
\mciteBstWouldAddEndPuncttrue
\mciteSetBstMidEndSepPunct{\mcitedefaultmidpunct}
{\mcitedefaultendpunct}{\mcitedefaultseppunct}\relax
\EndOfBibitem
\bibitem{phistar}
A.~Banfi {\em et~al.}, \ifthenelse{\boolean{articletitles}}{\emph{{Optimisation
  of variables for studying dilepton transverse momentum distributions at
  hadron colliders}},
  }{}\href{https://doi.org/10.1140/epjc/s10052-011-1600-y}{Eur.\ Phys.\ J.\
  \textbf{C71} (2011) 1600},
  \href{http://arxiv.org/abs/1009.1580}{{\normalfont\ttfamily
  arXiv:1009.1580}}\relax
\mciteBstWouldAddEndPuncttrue
\mciteSetBstMidEndSepPunct{\mcitedefaultmidpunct}
{\mcitedefaultendpunct}{\mcitedefaultseppunct}\relax
\EndOfBibitem
\bibitem{LHCb-DP-2014-001}
R.~Aaij {\em et~al.}, \ifthenelse{\boolean{articletitles}}{\emph{{Performance
  of the LHCb Vertex Locator}},
  }{}\href{https://doi.org/10.1088/1748-0221/9/09/P09007}{JINST \textbf{9}
  (2014) P09007}, \href{http://arxiv.org/abs/1405.7808}{{\normalfont\ttfamily
  arXiv:1405.7808}}\relax
\mciteBstWouldAddEndPuncttrue
\mciteSetBstMidEndSepPunct{\mcitedefaultmidpunct}
{\mcitedefaultendpunct}{\mcitedefaultseppunct}\relax
\EndOfBibitem
\bibitem{LHCb-DP-2017-001}
P.~d'Argent {\em et~al.}, \ifthenelse{\boolean{articletitles}}{\emph{{Improved
  performance of the LHCb Outer Tracker in LHC Run 2}},
  }{}\href{https://doi.org/10.1088/1748-0221/12/11/P11016}{JINST \textbf{12}
  (2017) P11016}, \href{http://arxiv.org/abs/1708.00819}{{\normalfont\ttfamily
  arXiv:1708.00819}}\relax
\mciteBstWouldAddEndPuncttrue
\mciteSetBstMidEndSepPunct{\mcitedefaultmidpunct}
{\mcitedefaultendpunct}{\mcitedefaultseppunct}\relax
\EndOfBibitem
\bibitem{LHCb-DP-2012-002}
A.~A. Alves~Jr.\ {\em et~al.},
  \ifthenelse{\boolean{articletitles}}{\emph{{Performance of the LHCb muon
  system}}, }{}\href{https://doi.org/10.1088/1748-0221/8/02/P02022}{JINST
  \textbf{8} (2013) P02022},
  \href{http://arxiv.org/abs/1211.1346}{{\normalfont\ttfamily
  arXiv:1211.1346}}\relax
\mciteBstWouldAddEndPuncttrue
\mciteSetBstMidEndSepPunct{\mcitedefaultmidpunct}
{\mcitedefaultendpunct}{\mcitedefaultseppunct}\relax
\EndOfBibitem
\bibitem{LHCb-DP-2012-004}
R.~Aaij {\em et~al.}, \ifthenelse{\boolean{articletitles}}{\emph{{The \lhcb
  trigger and its performance in 2011}},
  }{}\href{https://doi.org/10.1088/1748-0221/8/04/P04022}{JINST \textbf{8}
  (2013) P04022}, \href{http://arxiv.org/abs/1211.3055}{{\normalfont\ttfamily
  arXiv:1211.3055}}\relax
\mciteBstWouldAddEndPuncttrue
\mciteSetBstMidEndSepPunct{\mcitedefaultmidpunct}
{\mcitedefaultendpunct}{\mcitedefaultseppunct}\relax
\EndOfBibitem
\bibitem{Sjostrand:2007gs}
T.~Sj\"{o}strand, S.~Mrenna, and P.~Skands,
  \ifthenelse{\boolean{articletitles}}{\emph{{A brief introduction to PYTHIA
  8.1}}, }{}\href{https://doi.org/10.1016/j.cpc.2008.01.036}{Comput.\ Phys.\
  Commun.\  \textbf{178} (2008) 852},
  \href{http://arxiv.org/abs/0710.3820}{{\normalfont\ttfamily
  arXiv:0710.3820}}\relax
\mciteBstWouldAddEndPuncttrue
\mciteSetBstMidEndSepPunct{\mcitedefaultmidpunct}
{\mcitedefaultendpunct}{\mcitedefaultseppunct}\relax
\EndOfBibitem
\bibitem{Sjostrand:2006za}
T.~Sj\"{o}strand, S.~Mrenna, and P.~Skands,
  \ifthenelse{\boolean{articletitles}}{\emph{{PYTHIA 6.4 physics and manual}},
  }{}\href{https://doi.org/10.1088/1126-6708/2006/05/026}{JHEP \textbf{05}
  (2006) 026}, \href{http://arxiv.org/abs/hep-ph/0603175}{{\normalfont\ttfamily
  arXiv:hep-ph/0603175}}\relax
\mciteBstWouldAddEndPuncttrue
\mciteSetBstMidEndSepPunct{\mcitedefaultmidpunct}
{\mcitedefaultendpunct}{\mcitedefaultseppunct}\relax
\EndOfBibitem
\bibitem{LHCb-PROC-2010-056}
I.~Belyaev {\em et~al.}, \ifthenelse{\boolean{articletitles}}{\emph{{Handling
  of the generation of primary events in Gauss, the LHCb simulation
  framework}}, }{}\href{https://doi.org/10.1088/1742-6596/331/3/032047}{J.\
  Phys.\ Conf.\ Ser.\  \textbf{331} (2011) 032047}\relax
\mciteBstWouldAddEndPuncttrue
\mciteSetBstMidEndSepPunct{\mcitedefaultmidpunct}
{\mcitedefaultendpunct}{\mcitedefaultseppunct}\relax
\EndOfBibitem
\bibitem{Allison:2006ve}
Geant4 collaboration, J.~Allison {\em et~al.},
  \ifthenelse{\boolean{articletitles}}{\emph{{Geant4 developments and
  applications}}, }{}\href{https://doi.org/10.1109/TNS.2006.869826}{IEEE
  Trans.\ Nucl.\ Sci.\  \textbf{53} (2006) 270}\relax
\mciteBstWouldAddEndPuncttrue
\mciteSetBstMidEndSepPunct{\mcitedefaultmidpunct}
{\mcitedefaultendpunct}{\mcitedefaultseppunct}\relax
\EndOfBibitem
\bibitem{Agostinelli:2002hh}
Geant4 collaboration, S.~Agostinelli {\em et~al.},
  \ifthenelse{\boolean{articletitles}}{\emph{{Geant4: A simulation toolkit}},
  }{}\href{https://doi.org/10.1016/S0168-9002(03)01368-8}{Nucl.\ Instrum.\
  Meth.\  \textbf{A506} (2003) 250}\relax
\mciteBstWouldAddEndPuncttrue
\mciteSetBstMidEndSepPunct{\mcitedefaultmidpunct}
{\mcitedefaultendpunct}{\mcitedefaultseppunct}\relax
\EndOfBibitem
\bibitem{LHCb-PROC-2011-006}
M.~Clemencic {\em et~al.}, \ifthenelse{\boolean{articletitles}}{\emph{{The
  \lhcb simulation application, Gauss: Design, evolution and experience}},
  }{}\href{https://doi.org/10.1088/1742-6596/331/3/032023}{J.\ Phys.\ Conf.\
  Ser.\  \textbf{331} (2011) 032023}\relax
\mciteBstWouldAddEndPuncttrue
\mciteSetBstMidEndSepPunct{\mcitedefaultmidpunct}
{\mcitedefaultendpunct}{\mcitedefaultseppunct}\relax
\EndOfBibitem
\bibitem{LHCb-DP-2013-002}
LHCb collaboration, R.~Aaij {\em et~al.},
  \ifthenelse{\boolean{articletitles}}{\emph{{Measurement of the track
  reconstruction efficiency at LHCb}},
  }{}\href{https://doi.org/10.1088/1748-0221/10/02/P02007}{JINST \textbf{10}
  (2015) P02007}, \href{http://arxiv.org/abs/1408.1251}{{\normalfont\ttfamily
  arXiv:1408.1251}}\relax
\mciteBstWouldAddEndPuncttrue
\mciteSetBstMidEndSepPunct{\mcitedefaultmidpunct}
{\mcitedefaultendpunct}{\mcitedefaultseppunct}\relax
\EndOfBibitem
\bibitem{POWHEGewZ}
L.~Barze {\em et~al.}, \ifthenelse{\boolean{articletitles}}{\emph{{Neutral
  current Drell-Yan with combined QCD and electroweak corrections in the POWHEG
  BOX}}, }{}\href{https://doi.org/10.1140/epjc/s10052-013-2474-y}{Eur.\ Phys.\
  J.\  \textbf{C73} (2013) 2474},
  \href{http://arxiv.org/abs/1302.4606}{{\normalfont\ttfamily
  arXiv:1302.4606}}\relax
\mciteBstWouldAddEndPuncttrue
\mciteSetBstMidEndSepPunct{\mcitedefaultmidpunct}
{\mcitedefaultendpunct}{\mcitedefaultseppunct}\relax
\EndOfBibitem
\bibitem{powheg1}
P.~Nason, \ifthenelse{\boolean{articletitles}}{\emph{{A New method for
  combining NLO QCD with shower Monte Carlo algorithms}},
  }{}\href{https://doi.org/10.1088/1126-6708/2004/11/040}{JHEP \textbf{11}
  (2004) 040}, \href{http://arxiv.org/abs/hep-ph/0409146}{{\normalfont\ttfamily
  arXiv:hep-ph/0409146}}\relax
\mciteBstWouldAddEndPuncttrue
\mciteSetBstMidEndSepPunct{\mcitedefaultmidpunct}
{\mcitedefaultendpunct}{\mcitedefaultseppunct}\relax
\EndOfBibitem
\bibitem{powheg2}
S.~Frixione, P.~Nason, and C.~Oleari,
  \ifthenelse{\boolean{articletitles}}{\emph{{Matching NLO QCD computations
  with Parton Shower simulations: the POWHEG method}},
  }{}\href{https://doi.org/10.1088/1126-6708/2007/11/070}{JHEP \textbf{11}
  (2007) 070}, \href{http://arxiv.org/abs/0709.2092}{{\normalfont\ttfamily
  arXiv:0709.2092}}\relax
\mciteBstWouldAddEndPuncttrue
\mciteSetBstMidEndSepPunct{\mcitedefaultmidpunct}
{\mcitedefaultendpunct}{\mcitedefaultseppunct}\relax
\EndOfBibitem
\bibitem{powheg3}
S.~Alioli, P.~Nason, C.~Oleari, and E.~Re,
  \ifthenelse{\boolean{articletitles}}{\emph{{A general framework for
  implementing NLO calculations in shower Monte Carlo programs: the POWHEG
  BOX}}, }{}\href{https://doi.org/10.1007/JHEP06(2010)043}{JHEP \textbf{06}
  (2010) 043}, \href{http://arxiv.org/abs/1002.2581}{{\normalfont\ttfamily
  arXiv:1002.2581}}\relax
\mciteBstWouldAddEndPuncttrue
\mciteSetBstMidEndSepPunct{\mcitedefaultmidpunct}
{\mcitedefaultendpunct}{\mcitedefaultseppunct}\relax
\EndOfBibitem
\bibitem{Pythia8Main}
T.~Sj\"ostrand {\em et~al.}, \ifthenelse{\boolean{articletitles}}{\emph{{An
  introduction to PYTHIA 8.2}},
  }{}\href{https://doi.org/10.1016/j.cpc.2015.01.024}{Comput.\ Phys.\ Commun.\
  \textbf{191} (2015) 159},
  \href{http://arxiv.org/abs/1410.3012}{{\normalfont\ttfamily
  arXiv:1410.3012}}\relax
\mciteBstWouldAddEndPuncttrue
\mciteSetBstMidEndSepPunct{\mcitedefaultmidpunct}
{\mcitedefaultendpunct}{\mcitedefaultseppunct}\relax
\EndOfBibitem
\bibitem{LHCb-DP-2014-002}
LHCb collaboration, R.~Aaij {\em et~al.},
  \ifthenelse{\boolean{articletitles}}{\emph{{LHCb detector performance}},
  }{}\href{https://doi.org/10.1142/S0217751X15300227}{Int.\ J.\ Mod.\ Phys.\
  \textbf{A30} (2015) 1530022},
  \href{http://arxiv.org/abs/1412.6352}{{\normalfont\ttfamily
  arXiv:1412.6352}}\relax
\mciteBstWouldAddEndPuncttrue
\mciteSetBstMidEndSepPunct{\mcitedefaultmidpunct}
{\mcitedefaultendpunct}{\mcitedefaultseppunct}\relax
\EndOfBibitem
\bibitem{ALEPH:2005ab}
ALEPH, DELPHI, L3, OPAL, SLD, LEP Electroweak Working Group, SLD Electroweak
  Group, SLD Heavy Flavour Group, S.~Schael {\em et~al.},
  \ifthenelse{\boolean{articletitles}}{\emph{{Precision electroweak
  measurements on the $Z$ resonance}},
  }{}\href{https://doi.org/10.1016/j.physrep.2005.12.006}{Phys.\ Rept.\
  \textbf{427} (2006) 257},
  \href{http://arxiv.org/abs/hep-ex/0509008}{{\normalfont\ttfamily
  arXiv:hep-ex/0509008}}\relax
\mciteBstWouldAddEndPuncttrue
\mciteSetBstMidEndSepPunct{\mcitedefaultmidpunct}
{\mcitedefaultendpunct}{\mcitedefaultseppunct}\relax
\EndOfBibitem
\end{mcitethebibliography}
\newpage
\centerline
{\large\bf LHCb collaboration}
\begin
{flushleft}
\small
R.~Aaij$^{35}$\lhcborcid{0000-0003-0533-1952},
A.S.W.~Abdelmotteleb$^{54}$\lhcborcid{0000-0001-7905-0542},
C.~Abellan~Beteta$^{48}$,
F.~Abudin{\'e}n$^{54}$\lhcborcid{0000-0002-6737-3528},
T.~Ackernley$^{58}$\lhcborcid{0000-0002-5951-3498},
B.~Adeva$^{44}$\lhcborcid{0000-0001-9756-3712},
M.~Adinolfi$^{52}$\lhcborcid{0000-0002-1326-1264},
P.~Adlarson$^{78}$\lhcborcid{0000-0001-6280-3851},
C.~Agapopoulou$^{46}$\lhcborcid{0000-0002-2368-0147},
C.A.~Aidala$^{79}$\lhcborcid{0000-0001-9540-4988}
% ...

{\footnotesize \it

% ...
$^{35}$Nikhef National Institute for Subatomic Physics, Amsterdam, Netherlands\\
% ...
$^{44}$Instituto Galego de F{\'\i}sica de Altas Enerx{\'\i}as (IGFAE), Universidade de Santiago de Compostela, Santiago de Compostela, Spain\\
% ...
$^{46}$European Organization for Nuclear Research (CERN), Geneva, Switzerland\\
% ...
$^{48}$Physik-Institut, Universit{\"a}t Z{\"u}rich, Z{\"u}rich, Switzerland\\
% ...
$^{52}$H.H. Wills Physics Laboratory, University of Bristol, Bristol, United Kingdom\\
% ...
$^{54}$Department of Physics, University of Warwick, Coventry, United Kingdom\\
% ...
$^{57}$School of Physics and Astronomy, University of Glasgow, Glasgow, United Kingdom\\
$^{58}$Oliver Lodge Laboratory, University of Liverpool, Liverpool, United Kingdom\\
% ...
$^{66}$Syracuse University, Syracuse, NY, United States\\
% ...
$^{78}$Department of Physics and Astronomy, Uppsala University, Uppsala, Sweden, associated to $^{57}$\\
$^{79}$University of Michigan, Ann Arbor, MI, United States, associated to $^{66}$\\
% ...
}
\end{flushleft}
\end{document}